\documentclass[12pt]{article}
\pdfoutput=1
\usepackage{amsmath,amssymb,amsfonts,amsthm,bm,bbm,cancel,wasysym}
\usepackage{epsfig,graphics,graphicx,epstopdf,caption,subcaption}
\usepackage[numbers,sort&compress]{natbib}
\graphicspath{{Charts/}}
\usepackage{array,booktabs,colortbl,colordvi,multirow}
\usepackage{colordvi,color,xcolor}
\usepackage{hyperref}
\usepackage{rotating}
\usepackage{comment}
\usepackage{feynmf}

\begin{document}

\begin{center}

{\Large {\bf Gravitational freeze-in dark matter from Higgs Preheating}}\\

\vspace*{0.75cm}

{Ruopeng Zhang, Zixuan Xu and Sibo Zheng}

\vspace{0.5cm}
{Department of Physics, Chongqing University, Chongqing 401331, China}

\end{center}
\vspace{.5cm}

\begin{abstract}
\noindent 

Gravitational freeze-in is a mechanism to explain the observed dark matter relic density if dark matter neither couples to inflation nor to standard model sector.
In this work we study gravitational freeze-in dark matter production during Higgs preheating based on non-perturbative resonance.
Using reliable lattice method to handle this process, 
we show that tachyonic resonance is prohibited by strong back reaction due to Higgs self interaction needed to keep the positivity of potential during preheating, 
and parameter resonance is viable by tuning the Higgs self-interaction coupling to be small enough in ultraviolet energy scale.
We then derive the dark matter relic density under the context of Higgs preheating, 
and uncover a new dark matter parameter space with dark matter mass larger than inflaton mass, 
which arises from out-of-equilibrium Higgs annihilation.
Finally, we briefly remark the open question of testing gravitational dark matter.
\end{abstract}

\renewcommand{\thefootnote}{\arabic{footnote}}
\setcounter{footnote}{0}
\thispagestyle{empty}
\vfill
\newpage
\setcounter{page}{1}

\tableofcontents

\section{Introduction}
It is well known that dark matter (DM) should be a new particle beyond standard model (SM).
No direct evidences exist for DM except its gravitational effects on the evolution of Universe so far.
If we strictly stick to this clue DM may be only produced via gravitational portal,
where it is challenging to address the observed DM relic density reported by Planck \cite{Aghanim:2018eyx}.
Obviously, a gravitational DM differs from the so-called weakly interacting massive particle (WIMP) based upon on freeze-out mechanism under which WIMP is able to keep thermal equilibrium with SM thermal bath in early Universe. 
Rather, the gravitational DM production favors freeze-in mechanism. 

Recently, the gravitational freeze-in DM production has been developed in depth. 
It takes place in the circumstance of reheating \cite{Bassett:2005xm} after the end of inflation.
A general action of reheating takes the form 
\begin{eqnarray}{\label{action}}
S=\int d^{4}x \sqrt{-g}\left[\frac{M^{2}_{P}}{2}R+\mathcal{L}_{\phi}+\mathcal{L}_{\rm{SM}}+\mathcal{L}_{\chi}+\mathcal{L}_{\rm{int}}^{\rm{G}}+\mathcal{L}_{\rm{int}}^{\rm{non-G}}\right],
\end{eqnarray}
Here, $M_{P}=2.4\times 10^{18}$ GeV is the reduced Planck mass, $R$ is the Ricci scalar, 
$\rm{SM}$, $\phi$ and $\chi$ refer to the SM, inflaton and DM sector respectively,
$\mathcal{L}_{\rm{int}}^{\rm{G}}$ denotes the gravitational interactions among these sectors given by
\begin{eqnarray}{\label{int}}
\mathcal{L}_{\rm{int}}^{\rm{G}}=-\frac{h^{\mu\nu}}{M_{P}}\left[T_{\mu\nu}^{\phi}+T_{\mu\nu}^{\rm{SM}}+T_{\mu\nu}^{\rm{\chi}}\right],
\end{eqnarray}
with $h_{\mu\nu}$ the gravitational field and $T^{(j)}_{\mu\nu}$ the energy-momentum tensor of each sector $j$, and finally $\mathcal{L}_{\rm{int}}^{\rm{non-G}}$ represents the non-gravitational interactions among these sector.
Neglecting $\mathcal{L}_{\rm{int}}^{\rm{non}-G}$, 
the authors of refs.\cite{Markkanen:2015xuw,Fairbairn:2018bsw,Hashiba:2018tbu,Ema:2019yrd,Ahmed:2020fhc,Babichev:2020yeo,Karam:2020rpa,Ling:2021zlj,Mambrini:2021zpp, Barman:2021ugy, Clery:2021bwz, Haque:2021mab, Kaneta:2022gug,Ahmed:2022tfm,Klaric:2022qly,Barman:2022qgt,Lebedev:2022vwf}  have studied the gravitational freeze-in DM production via inflaton annihilation $\phi\phi\rightarrow h_{\mu\nu}\rightarrow\chi\chi$, either with the minimal or non-minimal coupling to the Ricci scalar.
However, the inflaton annihilation is not always the main source, e.g. in the circumstances where the non-gravitational interaction term $\mathcal{L}_{\rm{int}}^{\rm{non}-G}$ is present. This is the main topic of this work.

We will consider the gravitational freeze-in DM production in a realistic reheating by 
taking into account preheating \cite{Amin:2014eta} via the SM Higgs doublet $H$, 
which will be referred to as Higgs preheating.
For the Higgs preheating to occur, 
\begin{eqnarray}{\label{int}}
\mathcal{L}_{\rm{int}}^{\rm{non-G}}=\mathcal{L}_{\phi-H},
\end{eqnarray}
is an essential interaction to make sure that a transfer of the inflaton energy density $\rho_{\phi}$ to the Higgs energy density $\rho_{H}$ takes place by means of non-perturbative process. 
During this non-perturbative process,
the DM number density receives the contributions due to the inflaton condensate and the Higgs annihilation
\begin{eqnarray}{\label{totn}}
n_{\rm{\chi}}=n\left(\phi\phi\rightarrow h_{\mu\nu}\rightarrow\chi\chi\right)+n\left(HH\rightarrow h_{\mu\nu}\rightarrow\chi\chi\right),
\end{eqnarray} 
where the later one can easily dominate over the former one as a result of the non-perturbative nature e.g. in the situation with DM mass larger than inflaton mass. 
Another key feature of the Higgs annihilation contribution is that the produced relativistic Higgs particles during the Higgs preheating are out-of-equilibrium,
which differs from the thermalized SM particle annihilation contribution to the DM number density in the later reheating process \cite{Garny:2015sjg,Tang:2017hvq,Garny:2017kha,Chianese:2020yjo}. 
This difference has to be taken care of when one calculates the Higgs annihilation contribution to $n_{\chi}$ in eq.(\ref{totn}). 

Allowing other direct interactions between the DM and inflation/ SM sector in eq.(\ref{int}) can lead to additional non-gravitational DM production \cite{Bernal:2018hjm,Lebedev:2021tas,Garcia:2021iag,Garcia:2022vwm}, which is beyond the scope of this work.

The rest of the paper is organized as follows. 
In Sec.\ref{model} we introduce the model parameters in the Higgs preheating,
and discuss the allowed ranges by taking into account constraints from cosmological observations on the inflaton sector as well as collider data about the Higgs sector.
In Sec.\ref{pre} we present analytical formulae of Higgs preheating which will be divided into two cases namely tachyonic and parameter resonance, and use a publicly available lattice code to calculate relevant parameters such as $\rho_{\phi}$, $\rho_{H}$, $n_{\chi}$ etc all of which are variables of time.
Sec.\ref{GDM} is devoted to analyze the gravitational freeze-in DM production in the case of DM mass larger than the inflaton mass, which only arises from the Higgs annihilation during the preheating.
We firstly derive an analytical formula for the gravitational freeze-in DM relic density during the Higgs preheating in Sec.\ref{f},
then show the numerical results of DM parameter space in Sec.\ref{results}.
In Sec.\ref{s} we briefly discuss phenomenology of gravitational DM. 
Finally, we conclude in Sec.\ref{con}.

\section{The model}
\label{model}
We consider the following Lagrangian of inflaton and Higgs sector
\begin{eqnarray}
 \mathcal{L}_{\phi}&=& \frac{1}{2} \partial_{\mu}\phi\partial^{\mu}\phi -V(\phi), \label{Lin}\\
\mathcal{L}_{H}&=& \left(D_{\mu}H\right)^{\dag}D^{\mu}H -V(H), \label{LSM}\\
\mathcal{L}_{\phi-H}&=& 
 \left\{
\begin{array}{lcl}
-\sigma M_{P}\mid H\mid^{2}\phi, ~~~~~~~~~\rm{cubic}\\
-\kappa\mid H\mid^{2}\phi^{2}, ~~~~~~~~~~~~\rm{quartic}\\
\end{array}\right.
\label{Lint}
\end{eqnarray}
Explicitly, 
\begin{itemize}
\item For the inflaton sector we employ the $\alpha$-attractor T-model of inflation with  $V(\phi)=\lambda_{\phi} M_{P}^{4}\left[\sqrt{6}\tanh\left(\frac{\mid\phi\mid}{\sqrt{6}M_{P}}\right)\right]^{2}$ \cite{Kallosh:2013hoa,Kallosh:2013yoa}. 
Expanding $V(\phi)$ around the minimal value at $\phi=0$, we have $V(\phi)\simeq \frac{1}{2}m^{2}_{\phi}\phi^{2}$ with $m_{\phi}=\sqrt{2\lambda_{\phi}}M_{P}$.
\item For the SM Higgs doublet $H=\frac{1}{\sqrt{2}}(h^{+}_{1}+ih^{+}_{2}, h^{0}_{3}+ih^{0}_{4})^{T}$ with $h_i$ the four scalar components, 
and its self interaction reads as $V(H)\approx \lambda_{H}\mid H\mid^{4}$ where the small weak scale $\upsilon$ is safely neglected.
 \item For the interaction between these two sectors we consider either cubic or quartic interaction with  dimensionless coupling $\sigma$ (in units of Planck mass) and $\kappa$ respectively.
 The quartic and cubic interaction may allow a parameter \cite{Kofman:1994rk,Shtanov:1994ce,Kofman:1997yn} and tachyonic \cite{Felder:2000hj,Felder:2001kt} resonance respectively.  
 \end{itemize}

Along with the inflation and Higgs sector defined by eqs.(\ref{Lin})-(\ref{Lint}), 
we consider the DM sector being a Dirac fermion $\chi$ with Lagrangian 
\begin{eqnarray}
\mathcal{L}_{\chi}&=& i\bar{\chi}\gamma^{\mu}\partial_{\mu}\chi-m_{\chi}\bar{\chi}\chi \label{LDM}+\cdots,
\end{eqnarray}
with $m_{\chi}$ the DM mass, up to other interactions in this sector that are irrelevant about the gravitational production of $\chi$.
Note, if $\chi$ is instead a scalar degree of freedom, an additionally gravitational production of $\chi$ during inflation (in small momentum regions) has to be taken into account, see refs.\cite{Ling:2021zlj,Kaneta:2022gug}.

\subsection{Constraints}
Before we study the Higgs preheating let us discuss constraints on the model parameters therein.

\begin{itemize}
\item The inflation sector is constrained by the cosmological observations on cosmic microwave background (CMB), where the value of $\lambda_{\phi}$ has been fixed to be $\lambda_{\phi}\simeq 2.05\times 10^{-11}$at the pivot scale $k_{*}=0.05$ Mpc$^{-1}$  by Planck  \cite{Planck:2018jri}.
\item The Higgs sector is constrained by LHC, where the reported Higgs mass \cite{ATLAS:2012yve,CMS:2012qbp} has fixed the Higgs self-interaction coupling constant to be $\lambda_{H}\approx 0.13$ at the weak scale $\upsilon$. 
In the context of SM renormalization group equations, 
the value of $\lambda_H$ decreases when the renormalization scale is run from $\upsilon$ towards to $M_P$,
which causes the so-called issue of SM vacuum stability \cite{Elias-Miro:2011sqh}, as the sign of $\lambda_H$ flips in some intermediate mass scale between $\upsilon$ and $M_P$.
This issue can be evaded e.g. by an extension on the SM sector.
Depending on the extension the value of $\lambda_H$ in ultraviolet region close to $M_P$ varies in certain range.
Without losing generality we will consider 
\begin{eqnarray}\label{constraint1}
0 \leq \lambda_{H} \leq 0.13.
\end{eqnarray}
\item The interaction in Eq.(\ref{Lint}) is constrained by inflation and reheating. Firstly, during inflation where $\phi$ is of order a few times of $M_P$ the requirement of inflation dominated by $V(\phi)$ is trivially satisfied by a positive $\sigma$ or $\kappa$ as we assume here,
and the flatness of inflation potential is kept by imposing $\sigma^{2}<\lambda_{\phi}$ or $\kappa^{2}<\lambda_{\phi}$, which are obtained by taking into account the Higgs induced loop correction \cite{Bernal:2018hjm} to the inflation potential $V(\phi)$. 
Moreover, during preheating the requirement of positivity of the potential imposes the condition $\sigma<\sigma_{c}=2\sqrt{\lambda_{H}\lambda_{\phi}}$ \cite{Dufaux:2006ee} for the case of cubic interaction.
Collecting these constraints we have
\begin{eqnarray}\label{constraint2}
0 &<&\sigma<\sigma_{c}, \nonumber\\
0&<&\kappa< \sqrt{\lambda_{\phi}},
\end{eqnarray}
up to coefficients of order unity.
The constraints in eq.(\ref{constraint2}) may be not affected by the mentioned extension on the SM sector, see e.g. refs.
\cite{Lebedev:2012zw, Elias-Miro:2012eoi}.
\end{itemize}
To summarize, the Higgs preheating defined by eqs.(\ref{Lin})-(\ref{Lint}) contains two constrained model parameters composed of either $\{\lambda_{H}, \sigma\}$ or $\{\lambda_{H}, \kappa\}$ for the quartic and cubic interaction respectively.

\section{Higgs preheating}
\label{pre}
In this section we study the Higgs preheating after the end of inflation by means of non-perturbative parameter or tachyonic resonance. 
Unlike in a perturbative way such as the inflaton decay to SM particles where a set of Boltzmann equations can be used to analyze the evolution of relevant parameters,
one has to handle the non-perturbative preheating by a lattice method.

To prepare for the lattice calculation,
we begin with the equations of motion for $\phi$ and $H$:
\begin{eqnarray}\label{eom}
\left(\frac{d^{2}}{dt^{2}}+3H\frac{d}{dt}\right)\phi+V_{,\phi} &=& 0, \nonumber\\
\left(\frac{d^{2}}{dt^{2}}-\frac{\nabla^{2}}{a^{2}}+3H\frac{d}{dt}+m^{2}_{h,i}\right)h_{i}&=&0, 
\end{eqnarray}
where $V_{,\phi}\equiv\partial V/\partial \phi$, $h_{i}$ are the four real scalars in $H$, 
$a$ is the scale factor, ``$\nabla$" is derivative over space coordinates, 
$m^{2}_{h,i}$ is the effective mass squared varying with time due to the oscillation of $\phi$ during the preheating with
\begin{equation}{\label{mass}}
m^{2}_{h,i}\approx
 \left\{
\begin{array}{lcl}
\sigma M_{P}\phi+\lambda_{H}(2h^{2}_{i}+\sum_{i}h^{2}_{i}), ~~~~~~~~~\rm{cubic}\\
\kappa\phi^{2}+\lambda_{H}(2h^{2}_{i}+\sum_{i}h^{2}_{i}), ~~~~~~~~~~~~\rm{quartic}\\
\end{array}\right.
\end{equation}
and $H$ is the Hubble parameter given by
\begin{eqnarray}\label{H}
H^{2}\approx\frac{1}{3M^{2}_{P}}(\rho_{\phi}+\rho_{H}+\rho_{int}).
\end{eqnarray}
with $\rho_{H}=\sum_{i}\rho_{h_{i}}$ and $\rho_{int}$ the interactive energy.

As usual, it is more convenient to introduce the re-scaled scalar field $\mathcal{H}_{i}\equiv ah_{i}$ and
make use of Fourier transformation 
\begin{eqnarray}\label{Fourier}
\mathcal{H}_{i}(t,\mathbf{x})=\int \frac{d^{3}\mathbf{p}}{(2\pi)^{3/2}} e^{-i\mathbf{p}\cdot\mathbf{x}}\left[\tilde{h}_{i}(t)a_{p}+\tilde{h}^{*}_{i}(t)a^{\dag}_{-p}\right].
\end{eqnarray}
where $\mathbf{p}$ is the comoving momentum with $p=\mid\mathbf{p}\mid$. 
Using eq.(\ref{Fourier}) one rewrites the equation of motion of $h_i$ in eq.(\ref{eom}) 
\begin{eqnarray}\label{eomm}
\tilde{h}_{i}''+\omega^{2}_{p}\tilde{h}_{i}=0
\end{eqnarray}
where ``prime" is derivative over the conformal time $\tau$ and
\begin{eqnarray}\label{omega}
\omega^{2}_{p}(t)=p^{2}-\frac{a''}{a}+a^{2}m^{2}_{\rm{h},i}.
\end{eqnarray}
With respect to each comoving momentum mode $p$, the particle occupation number is
\begin{eqnarray}\label{n}
n_{p}(t)=\frac{1}{2\omega_{p}}\mid \omega_{p}\tilde{h}_{i}-i\tilde{h}'_{i}\mid^{2},
\end{eqnarray}
while the energy reads as
\begin{eqnarray}\label{E}
E_{p}(t)=\omega_{p}(n_{p}+\frac{1}{2})=\frac{1}{2}\mid \tilde{h}_{i}'\mid^{2}+\omega^{2}_{p}\mid \tilde{h}_{i}\mid^{2}.
\end{eqnarray}
According to eq.(\ref{n}) and  eq.(\ref{E}) one obtains the number and energy density of individual $h_i$ \cite{Kofman:1997yn}:
\begin{eqnarray}\label{ed}
n_{h_{i}}(t)&=&\frac{1}{(2\pi)^{2}}\left(\frac{a_{\rm{end}}}{a}\right)^{3}\int d^{3}\mathbf{p}~n_{p}(t), \nonumber\\
\rho_{h_{i}}(t)&=&\frac{1}{(2\pi)^{2}}\left(\frac{a_{\rm{end}}}{a}\right)^{4}\int d^{3}\mathbf{p}~n_{p}(t)\omega_{p}(t).
\end{eqnarray}
A more practical use of the number density in lattice calculation is the phase-space distribution function (PDF) given by
\begin{eqnarray}\label{pdf}
f(p,t)\equiv n_{p}(t)
\end{eqnarray}

With all necessary equations at hand, we are ready to numerically solve them in terms of the publicly available lattice code CosmoLattice \cite{Figueroa:2020rrl, Figueroa:2021yhd}.
Let us clarify that 
\begin{itemize}
\item The use of this lattice calculation to the Higgs preheating is different from those aimed to study DM preheating \cite{Bernal:2018hjm,Lebedev:2021tas,Garcia:2021iag,Garcia:2022vwm} where DM plays the role of $H$.
\item The Higgs preheating cannot be accounted for by four copies of singlet scalar preheating\footnote{For the early studies on the singlet scalar preheating see refs.\cite{Kofman:1994rk,Shtanov:1994ce,Kofman:1997yn,Traschen:1990sw,Kaiser:1995fb,Khlebnikov:1996mc}.} due to the non-perturbative nature of preheating.
\item More importantly, the singlet scalar preheating cannot mimic the Higgs preheating, 
as the Higgs self interaction has been experimentally constrained at the weak scale. 
\end{itemize}
We explicitly use the following set of initial conditions:
\begin{eqnarray}\label{initial}
\rho_{\phi}\mid_{\rm{end}}&=&\frac{3}{2}V_{\rm{end}}=7.1\times 10^{62}\rm{GeV}^{4},\nonumber\\
\rho_{H}\mid_{\rm{end}}&=&0, \nonumber\\
\phi\mid_{\rm{end}}&=&0.84~M_{P},
\end{eqnarray}
together with the initial conditions of Bunch-Davies vacuum for $\tilde{h}_{i}$, 
where the subscript ``end" denotes the end time of inflation.
Other initial conditions would be mentioned elsewhere  if necessary.

\begin{figure}
\centering
\includegraphics[width=8cm,height=8cm]{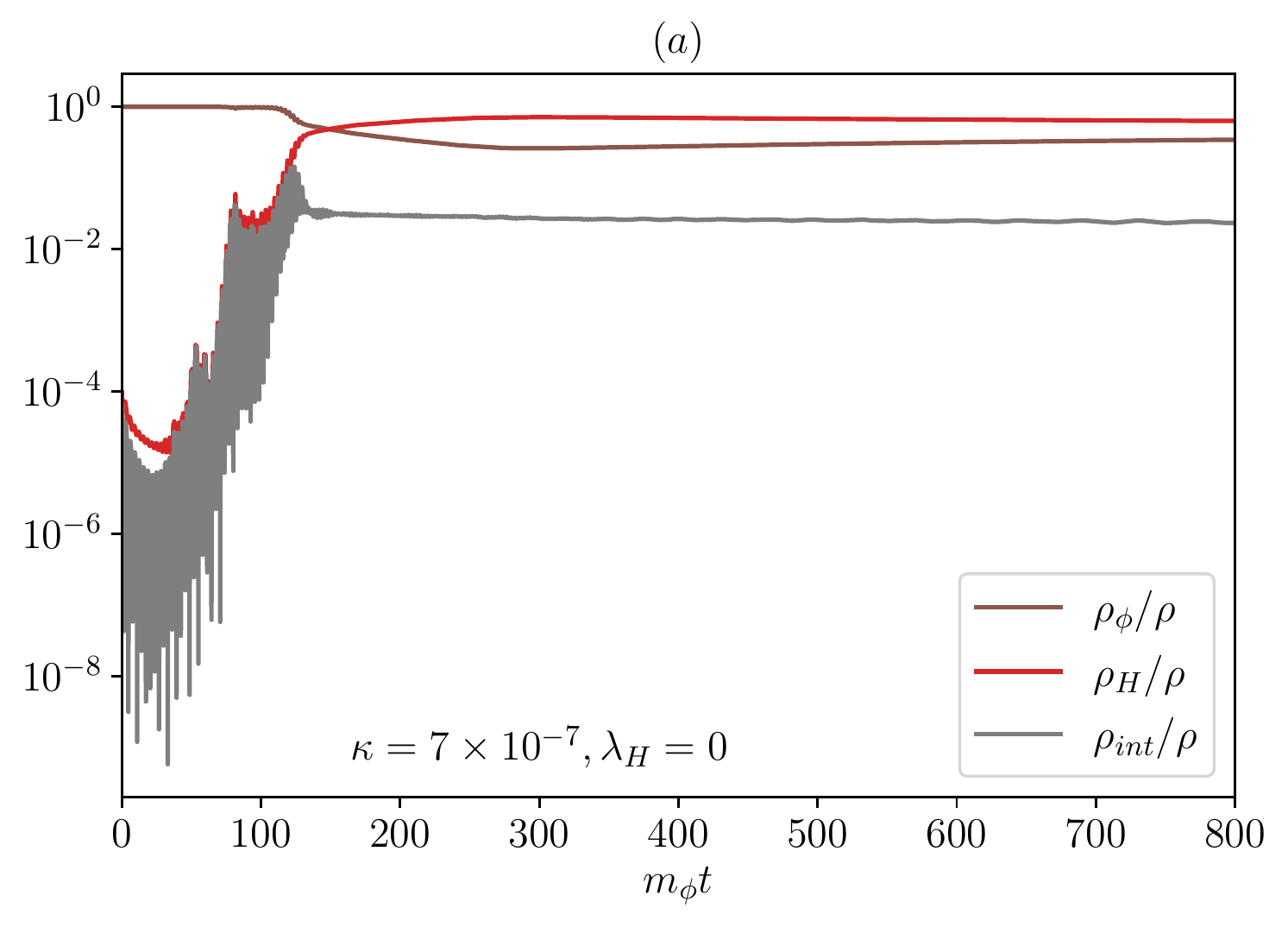}
\includegraphics[width=8cm,height=8cm]{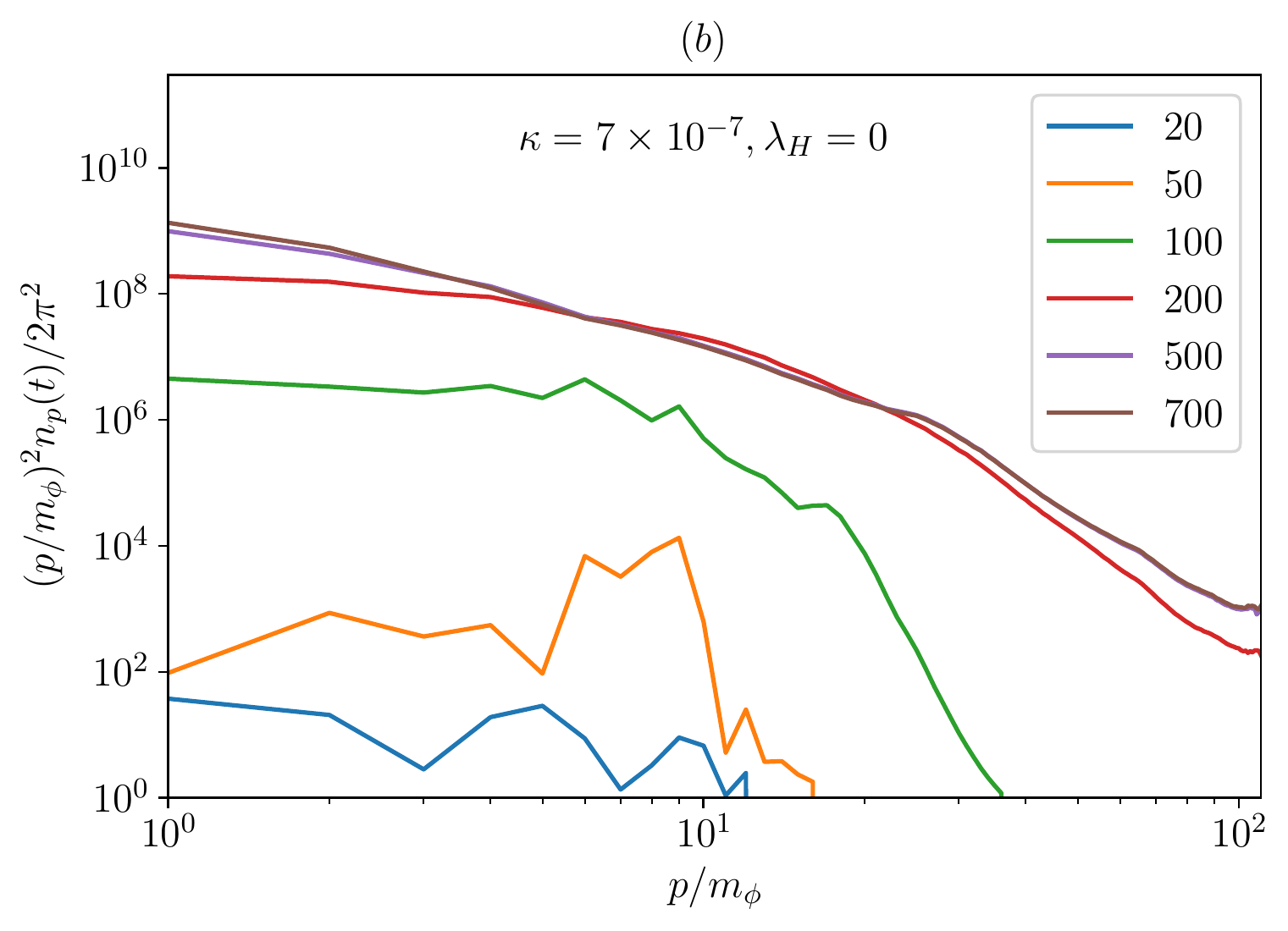}
\centering
\caption{Parameter resonance for $\lambda_{H}=0$ and $\kappa=7\times10^{-7}$. (a): evolution of energy densities $\rho_{\phi}$, $\rho_{H}$ and $\rho_{\rm{int}}$ normalized to the total energy density $\rho=\rho_{\phi}+\rho_{H}+\rho_{\rm{int}}$ with its initial value given by eq.(\ref{initial}). (b): The occupation number of Higgs scalar $n_{p}(t)$ which evolves with both momentum $p$ and time $t$, where numbers referring to colors are the values of $m_{\phi}t$.}
\label{r1}
\end{figure}

\subsection{Quartic interaction}
According to the constraints in eqs.(\ref{constraint1})-(\ref{constraint2}), 
we firstly use a set of parameter choices with $\lambda_{H}=0$ and $\kappa=7\times10^{-7}$ to illustrate the viability of Higgs preheating via parameter resonance.
Fig.\ref{r1}(a) shows the evolution of energy densities $\rho_{\phi}$, $\rho_{H}$ and $\rho_{\rm{int}}$ as function of time in units of $m_{\phi}$, all of which are normalized to the total energy density $\rho=\rho_{\phi}+\rho_{H}+\rho_{\rm{int}}$ whose initial value is given in eq.(\ref{initial}).
One sees the non-perturbative growth of $\rho_{H}$, which approaches to nearly $\sim 70\%$ of the total energy density at the time earlier than $\sim 200/m_{\phi}$.
Fig.\ref{r1}(b) shows the occupation number of Higgs scalar $n_{p}(t)$ defined in eq.(\ref{n}) as function of both momentum $p$ and time.
It explosively grows and is rapidly stabilized after the time later than $\sim 200/m_{\phi}$.
We will use this numerical result of $n_{p}(t)$ to derive the Higgs scalar PDF defined in eq.(\ref{pdf}) for calculating the gravitational freeze-in DM relic density.
Note, the pattern of $n_{p}(t)$ indicates that the Higgs scalar PDF is different from the Bose-Einstein distribution.  

\begin{figure}
\centering
\includegraphics[width=8cm,height=8cm]{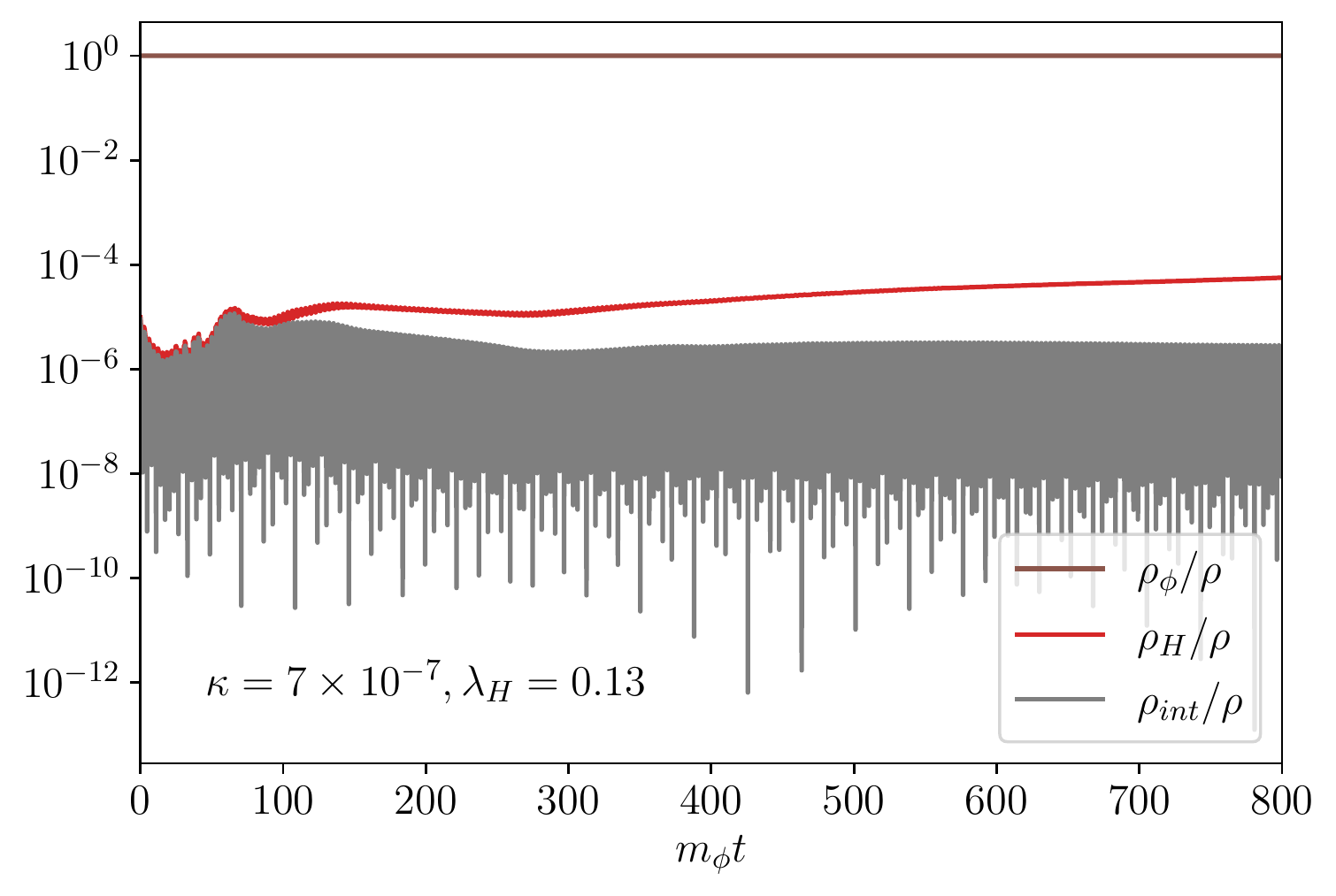}
\centering
\caption{Same as fig.\ref{r1}(a) with $\lambda_{H}=0.13$ and $\kappa=7\times10^{-7}$ instead, 
which indicates that the Higgs preheating via parameter resonance is spoiled by strong back reaction due to the Higgs self interaction.}
\label{r2}
\end{figure}

In order to illustrate that the above Higgs preheating via the mechanism of parameter resonance may be spoiled by the Higgs self interaction, we take a set of parameter choices with $\lambda_{H}=0.13$ and $\kappa=7\times10^{-7}$ compared to the previous ones. 
Similar to fig.\ref{r1}(a) we show in fig.\ref{r2} the evolution of relevant energy densities,
where the energy transfer from $\rho_{\phi}$ to $\rho_H$ is no longer sufficient.
This is caused by strong back reaction\footnote{Similar back reaction was reported by ref.\cite{Bernal:2018hjm} in the context of DM preheating.} due to the Higgs self interaction. 
Just like the back reaction \cite{Kofman:1997yn} on the inflaton due to the loop effect  of $\left<hh\right>$ induced by the quartic interaction,
there is a similar back reaction on the Higgs scalar fields due to the Higgs self interaction.
Following Hartree approximation \cite{Kofman:1997yn}, 
one finds that this Higgs self interaction alters the Higgs effective mass in eq.(\ref{mass}) as $m^{2}_{h_{i}}=\kappa \phi^{2}+C\lambda_{H}n_{p}(t)/\mid\phi\mid$ with $C$ a complex coefficient.
Whenever the oscillating $\phi$ crosses the zero point, 
the Higgs effective mass no longer vanishes but is taken over by the Higgs self-interaction induced mass term. 
If this mass is heavy enough it can dramatically suppress an explosive production of Higgs particles.
To make the back reaction under control, we can tune the value of $\lambda_H$  to be negligible in the ultraviolet energy region by the extension on the SM sector.
For the values of $\kappa\sim 10^{-7}-10^{-6}$ under consideration, 
the lattice analysis shows that the parameter resonance is maintained by $\lambda_{H}\leq 10^{-5}$.

\subsection{Cubic interaction}
\begin{figure}
\centering
\includegraphics[width=8cm,height=8cm]{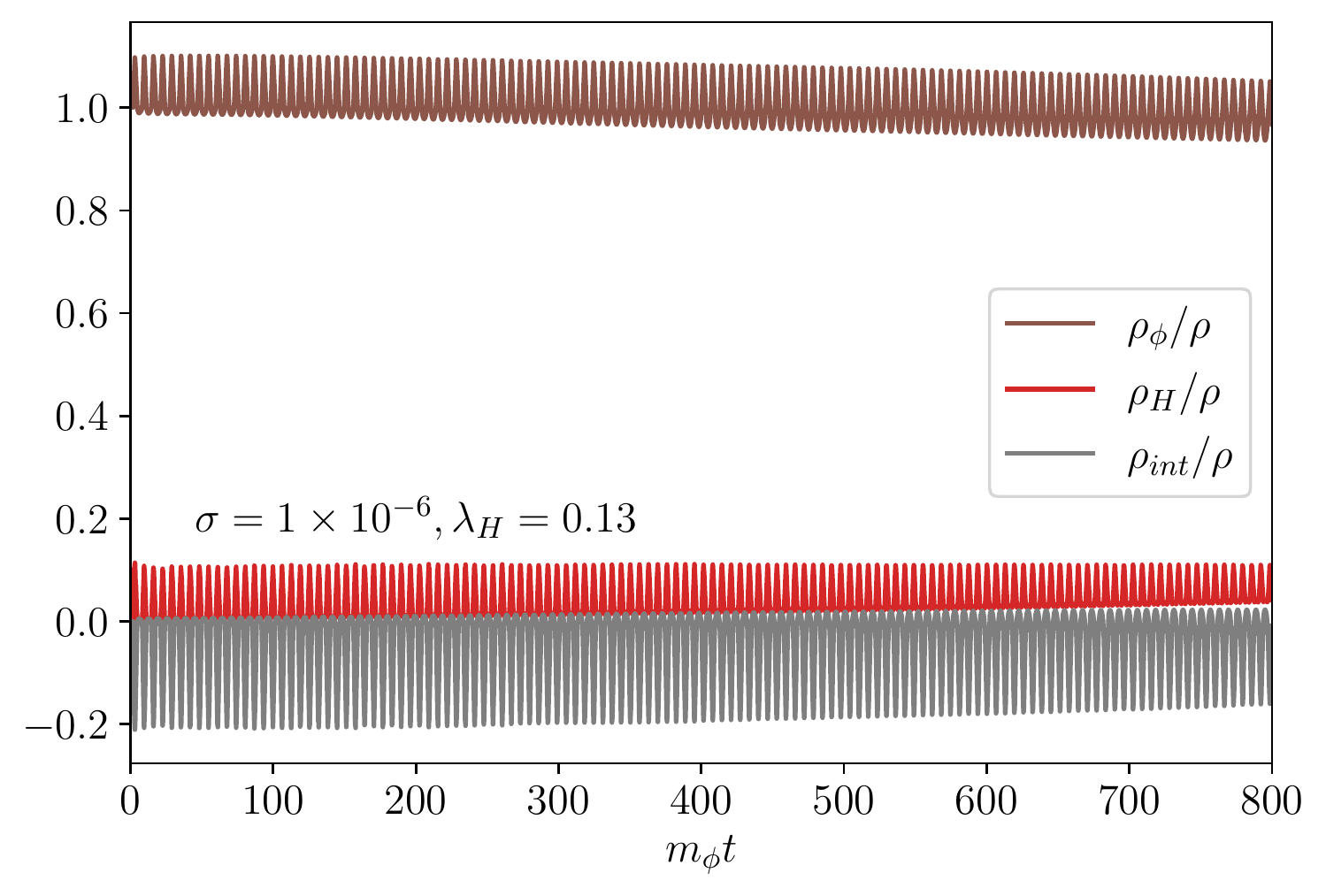}
\centering
\caption{An illustration to show the exclusion of Higgs preheating via tachyonic resonance for $\sigma=1.0\times 10^{-6}$ which is close to its upper bound $\sigma_{c}$. Similar results are verified by other choices on smaller values of $\sigma$.}
\label{t}
\end{figure}

The Higgs preheating may be alternatively realized via tachyonic resonance due to the cubic interaction. 
Since tachyonic resonance favors a large value of $\sigma$, 
for illustration we show the relevant energy transfer for $\sigma=1.0\times 10^{-6}$ close to the upper bound $\sigma_{c}\approx 3\times 10^{-6}$ in fig.\ref{t}.
It reveals that the energy transfer becomes insufficient due to strong back reaction arising from the Higgs self interaction, 
similar to what occurs in fig.\ref{r2}.
One may wonder whether the tachyonic resonance revives by turning off the Higgs self interactions with $\lambda_{H}\rightarrow 0$.
However, adopting a smaller $\lambda_H$ leads to a smaller upper bound $\sigma_{c}$ as seen in eq.(\ref{constraint2}), 
which disfavors the tachyonic resonance again.
By repeating other choices on smaller values of $\sigma$, 
we have verified that the Higgs preheating via tachyonic resonance is indeed implausible.

Before closing the study of Higgs preheating, 
let us summarize the main result of this section - the idea of Higgs preheating can be only achieved by the mechanism of parameter resonance with the Higgs self interaction small enough in the ultraviolet energy scale.

\section{Gravitational freeze-in dark matter}
\label{GDM}

\subsection{Relic density}
\label{f}
The gravitational freeze-in DM production occurs during both the preheating and the later reheating.
In the former epoch the energy density $\rho_{\phi}$ is transferred to $\rho_{H}$ via the parameter resonance where both inflaton and Higgs annihilation contribute to the DM relic density,
whereas in the later period  assumed to be in broken electroweak phase $\rho_{H}$ is transferred to $\rho_{\rm{SM}}$ of SM particles via the Higgs decay where the SM particle annihilations further contribute to the DM relic density.

Let us firstly address the gravitional freeze-in production during the preheating.
Despite the evolutions of $\rho_{\phi}$ and $\rho_{H}$ being non-perturbative,
it is still valid to to analyze the gravitational freeze-in production via the Boltzmann equation 
\begin{eqnarray}\label{Boltn}
\dot{n}_{\chi}+3Hn_{\chi}&=&R_{\phi\phi\rightarrow \chi\chi}(t)+R_{HH\rightarrow \chi\chi}(t),
\end{eqnarray}
where $R_{\phi\phi\rightarrow \chi\chi}$ and  $R_{HH\rightarrow \chi\chi}$ is the inflaton condensate and the Higgs annihilation rate respectively. 

Calculating $R_{\phi\phi\rightarrow \chi\chi}$ in eq.(\ref{Boltn}) is simplified by the fact that $\phi$ is an excitation out of vacuum rather than a real particle. In this case \cite{Clery:2021bwz,Bernal:2018qlk}
\begin{eqnarray}\label{Ra}
R_{\phi\phi\rightarrow \chi\chi}(t)&=&\frac{1}{128\pi}\frac{\rho^{2}_{\phi}}{M^{4}_{P}}\left(1+\frac{m^{2}_{\chi}}{2m^{2}_{\phi}}\right)^{2}\sqrt{1-\frac{m^{2}_{\chi}}{m^{2}_{\phi}}},
\end{eqnarray}
where the dependence of $R_{\phi\phi\rightarrow \chi\chi}$ on $t$ comes from that of $\rho_{\phi}$.
Note, this formula was originally derived by an envelope approximation.

Calculating $R_{HH\rightarrow \chi\chi}$ in eq.(\ref{Boltn}) follows the Higgs annihilation rate weighted by the Higgs distribution function and integrated over relevant momentum space
\begin{eqnarray}\label{Rb1}
R_{HH\rightarrow \chi\chi}(t)\approx
\frac{4}{512\pi^{5}}\int f(E_{1},t)f(E_{2},t)E_{1}dE_{1}E_{2}dE_{2}d\cos\theta_{12}d\cos\theta_{13}\mid\mathcal{M}\mid_{12\rightarrow 34}^{2},
\end{eqnarray}
in terms of an appropriate treatment on graviton propagator,
where the factor ``4" accounts for the four scalar degrees of freedom in $H$,
$f$ is the Higgs PDF give by eq.(\ref{pdf}),  the subscript ``1" and ``2" refer to the two initial relativistic Higgs scalars with their energies and momenta \footnote{The physical momenta in eq.(\ref{Rb1}) are written by the corresponding comoving momenta by multiplying a factor $(a_{\rm{end}}/a)$.
As a result, the boundary condition of the integration in eq.(\ref{Rb1}) becomes 
\begin{eqnarray}{\label{boundary}}
p_{1}+p_{2}\geq 2m_{\chi}\left(\frac{a}{a_{\rm{end}}}\right).
\end{eqnarray}} satisfying $E_{1}=p_{1}$ and $E_{2}=p_{2}$ respectively while ``3" and ``4" are the two final state DM particles,  
$\theta_{13}$ and $\theta_{12}$ are two angles defined as $s=2E_{1}E_{2}(1-\cos\theta_{12})$ and $t=s(\cos\theta_{13}-1)/2$ with $s$ and $t$ two Mandelstam variables, 
and the squared amplitude is given by \cite{Clery:2021bwz} 
\begin{eqnarray}\label{M}
\mid\mathcal{M}\mid_{12\rightarrow 34}^{2}=\frac{-t(s+t)(s+2t)^{2}}{4M^{4}_{P}s^{2}}.
\end{eqnarray}
It is critical to note that the Higgs annihilation rate in eq.(\ref{Rb1}) differs from those of the earlier studies in refs.\cite{Clery:2021bwz,Bernal:2018qlk},
as the produced Higgs particles  during the preheating are out-of-equilibrium as shown by fig.\ref{r1}.

For the gravitational freeze-in production during the reheating, 
we define the end time of preheating and reheating as $t_{p}$ and $t_r$ respectively.
A natural guess of these time scales is $t_{p}\sim 1/m_{\phi}$ and $t_{r}\sim 1/\Gamma_{H}$ with $\Gamma_{H}$ the Higgs decay width.
In the reheating where the PDFs of the SM particles in eq.(\ref{Boltn}) are understood as the thermal-equilibrium distributions,
the SM annihilation rate is given by \cite{Garny:2017kha,Clery:2021bwz} 
\begin{eqnarray}\label{Rb2}
R_{\rm{SMSM}\rightarrow \chi\chi}\approx\sum_{K=\{0,1/2,1\}}N_{K}R^{(K)}_{\rm{SMSM}}\approx 0.017\frac{T_{\rm{r}}^{8}}{M^{4}_{P}},
\end{eqnarray}
where $N_{K}=(4,45,12)$ are the numbers of each SM species of spin and $T_{\rm{r}}$ is the temperature of the SM thermal bath during the reheating with
\begin{eqnarray}\label{Tr}
\rho_{\rm{SM}}=\frac{\pi^{2}}{30}g_{\rm{r}}(T_{\rm{r}})T_{\rm{r}}^{4},
\end{eqnarray}
with $g_{\rm{r}}\simeq 427/4$ the effective number of relativistic degrees of freedom of the SM thermal bath.
Since the produced Higgs particles and the SM thermal bath are both relativistic, we have $\rho_{\rm{SM}}=(a_{\rm{r}}/a_{\rm{p}})^{-4}\rho_{H}\approx (t_{\rm{r}}/t_{\rm{p}})^{-2}\rho_{H}$  with $a_{\rm{p}}$ and $a_{\rm{r}}$ the scale factor at the end of preheating and reheating respectively.
Plugging the value of $\rho_{\rm{SM}}$ into eq.(\ref{Tr}) by using $\rho_{H}\sim 10^{-5}m^{2}_{\phi}M^{2}_{P}$ one obtains $T_{\rm{r}}\sim 0.1\sqrt{\Gamma_{H}M_{P}}\sim 10^{6}$ GeV.
Compared with $R_{HH}$ in eq.(\ref{Rb1}),
$R_{\rm{SMSM}\rightarrow \chi\chi}$ in eq.(\ref{Rb2}) is much smaller. 

Therefore, with the SM annihilation contribution during the reheating safely neglected, 
the gravitational freeze-in DM relic density is mainly produced by 
the inflaton annihilation in the DM mass range with $m_{\chi}<m_{\phi}$ and the Higgs annihilation in the DM mass range with $m_{\chi}>m_{\phi}$ during the preheating, respectively. 

To derive the DM relic density in the DM mass region with $m_{\chi}>m_{\phi}$ which is of special interest to us,
we utilize the following formula in terms of the reheating parameters \cite{Clery:2021bwz}
\begin{eqnarray}\label{rf}
\Omega_{\chi}h^{2}=1.6\times 10^{8}\cdot\left(\frac{g_{0}}{g_{\rm{r}}}\right)\frac{n_{\chi,\rm{r}}}{T^{3}_{\rm{r}}}\left(\frac{m_{\chi}}{1~\rm{GeV}}\right),
\end{eqnarray}
where the subscript ``r" denotes the reheating with $g_{0}=43/11$.
Eq.(\ref{rf}) enables us to write the DM relic density in terms of the preheating parameters as
\begin{eqnarray}\label{pf}
\Omega_{\chi}h^{2}=1.6\times 10^{8}\cdot\left(\frac{g_{0}}{g_{\rm{r}}}\right)\frac{n_{\chi}}{\rho_{H}^{3/4}}\left(\frac{m_{\chi}}{1~\rm{GeV}}\right),
\end{eqnarray}
where $n_{\chi}$ is derived from eq.(\ref{Boltn}) without $R_{\phi\phi\rightarrow \chi\chi}$.

\subsection{Dark matter parameter space}
\label{results}

\begin{figure}
\centering
\includegraphics[width=8cm,height=8cm]{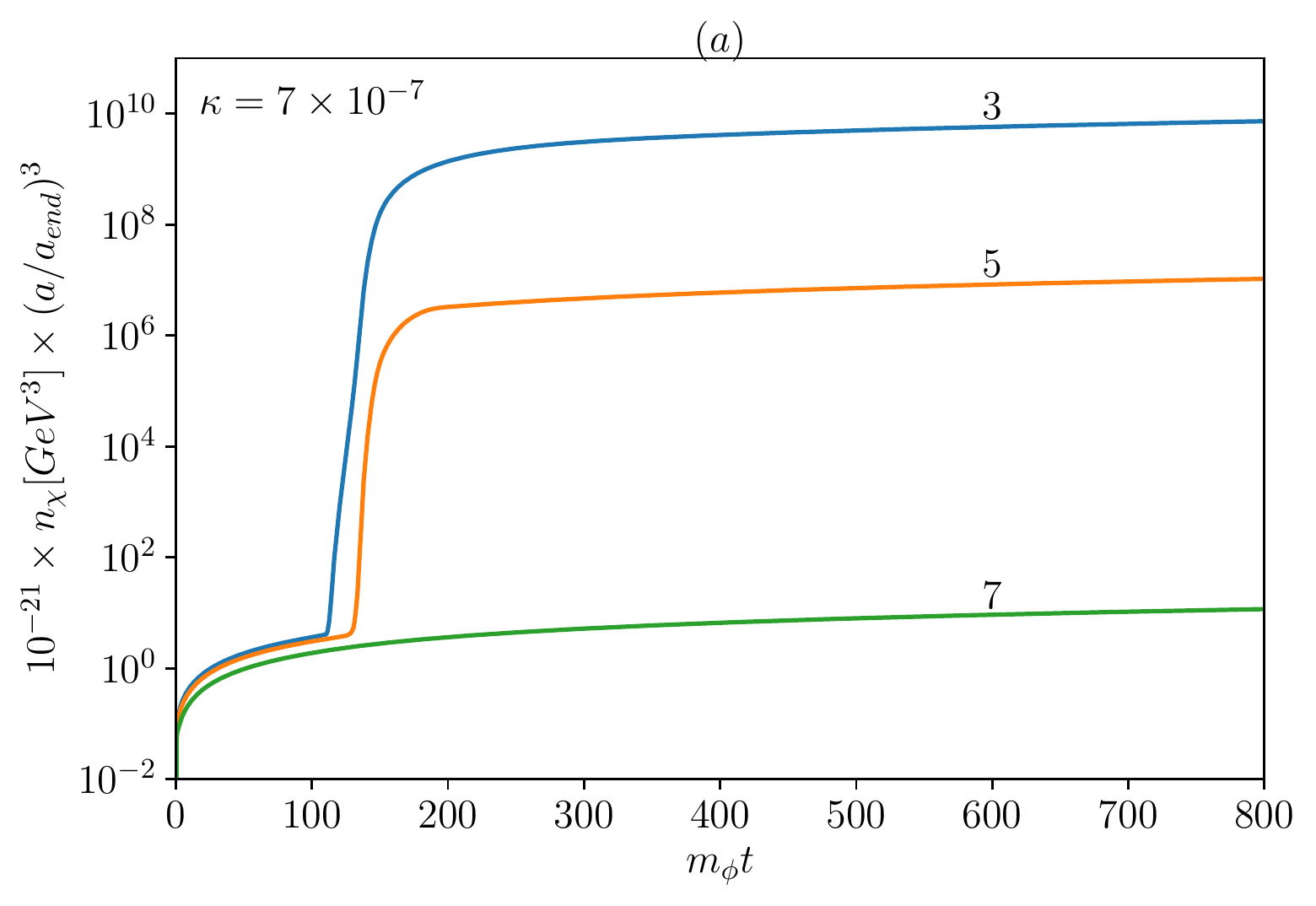}
\includegraphics[width=8cm,height=8cm]{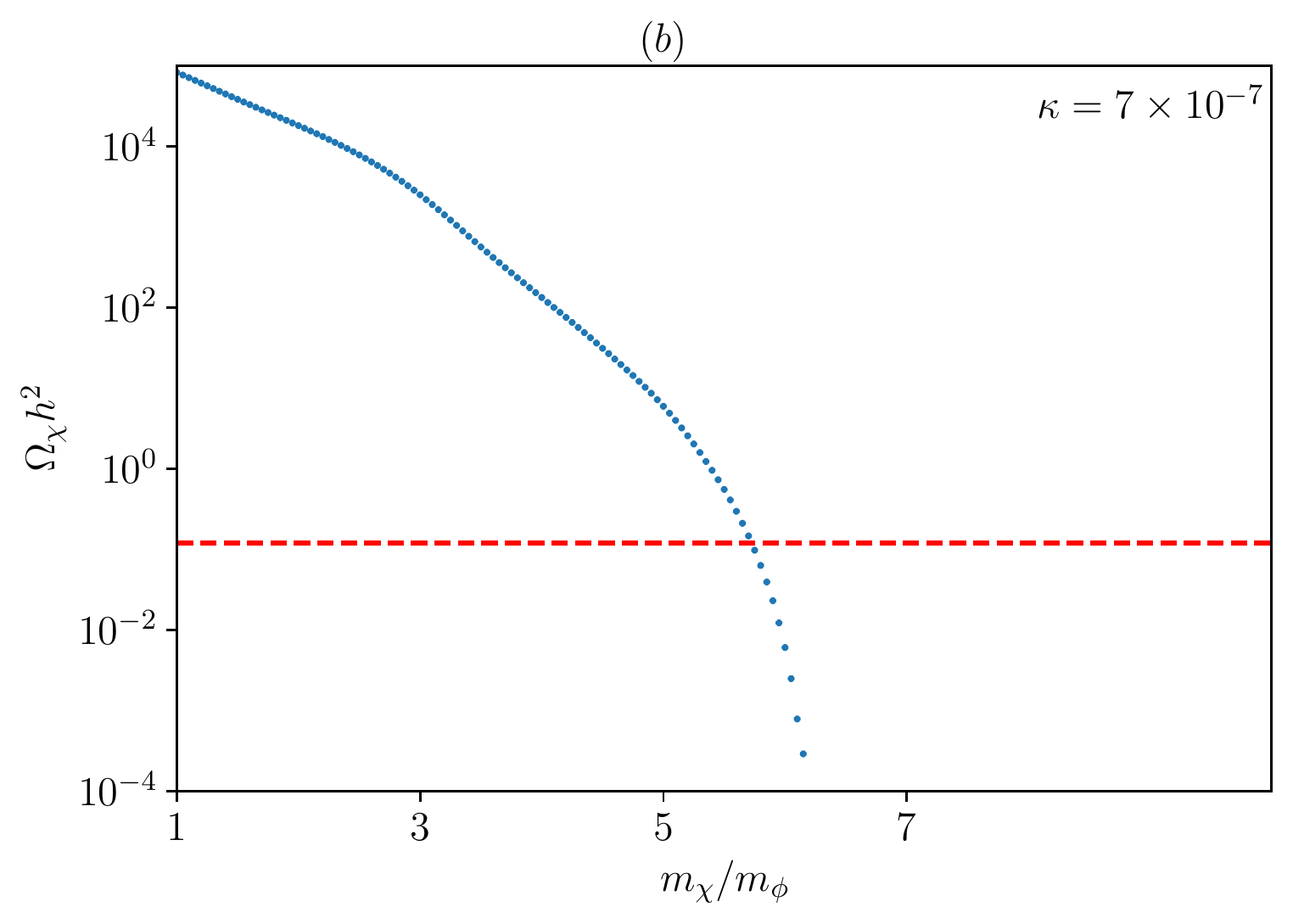}
\centering
\caption{Parameters in eq.(\ref{pf}) for a fixed value $\kappa=7\times10^{-7}$. Left: numerical values of the comoving number density $n_{\chi}a^{3}$ for various ratios of $m_{\chi}/m_{\phi}=\{3,5,7\}$. Right: gravitational freeze-in DM relic density in the DM mass range 
$m_{\chi}\geq m_{\phi}$.}
\label{fixed}
\end{figure}

\begin{figure}
\centering
\includegraphics[width=15cm,height=8cm]{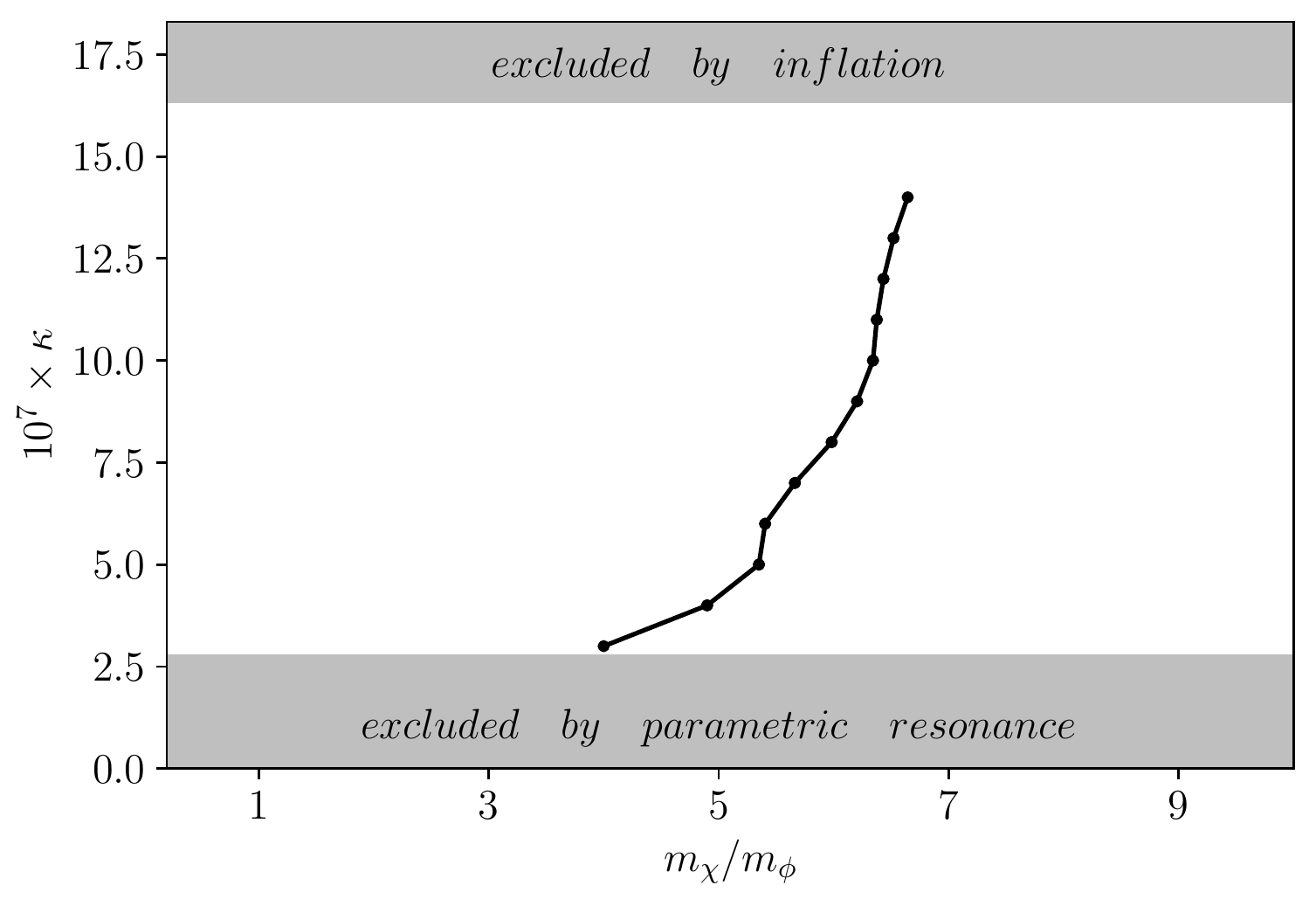}
\centering
\caption{DM parameter space with respect to $\Omega_{\chi}h^{2}=0.12\pm 0.01$ in the DM mass range with $m_{\chi}\geq m_{\phi}$. The gray shaded regions are excluded either by the constraint on inflation in eq.(\ref{constraint2}) or the adopted criteria of parameter resonance with $\rho_{H}\geq \rho_{\phi}$ at the end of preheating.}
\label{ps}
\end{figure}

Fig.\ref{fixed} shows the numerical results for $\kappa=7\times 10^{-7}$ and $\lambda_{H}=0$.
In the fig.\ref{fixed}(a) we give the evolution of $n_{\chi}$ as time $t$ for a few representative values of $m_{\chi}$ by numerically solving eq.(\ref{Boltn}).
The values of $n_{\chi}$ are stabilized within the adopted time scale, 
implying the estimate on DM relic density accurate enough.
In the fig.\ref{fixed}(b) we show the gravitational freeze-in DM relic density in the DM mass range 
$m_{\chi}\geq m_{\phi}$ for the same value of $\kappa=7\times 10^{-7}$,
where the red dotted line refers to the observed DM relic density reported by Planck.
It is satisfied by the two crossing points between the two curve lines.
The values of $\Omega_{\chi}h^{2}$ increase (decrease) by choosing a larger (smaller) $\kappa$,
suggesting the blue curve line to go up (down).
This trend will help us understand the DM parameter space as shown in fig.\ref{ps}.

Let us clarify that the threshold value of $m_{\chi}\leq 110~m_{\phi}$ is determined by our lattice calculation which corresponds to a lattice size $N=128^{3}$ and $k_{IR}=1$ with $k_{IR}$ the minimal infrared cut-off for the reciprocal lattice.
To uplift this threshold value, one has to take a higher lattice size.
For instance, the lattice size with $N=256^{3}$ gives a threshold value of $m_{\chi}\leq 220~m_{\phi}$ but with a cost of time increased by one order.

Instead of fixing $\kappa$, we now repeat the previous process for the allowed range of $\kappa$ as previously carried out. 
Fig.\ref{ps} shows the DM parameter space with respect to $\Omega_{\chi}h^{2}=0.12\pm 0.01$, 
where the gray shaded regions are excluded either by the constraint on inflation in eq.(\ref{constraint2}) or the adopted criteria of parameter resonance with $\rho_{H}\geq \rho_{\phi}$ at the end of preheating. 
These constraints narrow down the allowed value of $\kappa$ to be in the small range of $\sim (2-16)\times 10^{-7}$. 
The pattern of DM parameter space in fig.\ref{ps} can be inferred from the trend reflected by fig.\ref{fixed}.

\section{Phenomenology of gravitational dark matter}
\label{s}
While the idea of gravitational freeze-in DM production in the DM mass region with $m_{\chi}\geq m_{\phi}$ 
can be indeed achieved via the Higgs preheating,
how to test gravitational DM is an open question,
as neither DM direct detections nor colliders have a role to play.
Only cosmological or astrophysical experiments may be relevant.
In the following we briefly discuss this point in view of completeness.

We firstly consider DM annihilation into SM neutrinos and photons via the gravitational portal. 
The cross sections read as
\begin{eqnarray}
\sigma\left(\bar{\chi}\chi\rightarrow XX\right)\upsilon_{\rm{rel}}&\approx&\frac{1}{64\pi}\frac{m^{2}_{\chi}}{M^{4}_{P}} \times
 \left\{
\begin{array}{lcl}
\frac{1}{2}, ~~~~~~~~~~~~X=\nu\\
\frac{15}{8}, ~~~~~~~~~~~X=\gamma\\
\end{array}\right.
\label{ann}
\end{eqnarray}
where $\upsilon_{\rm{rel}}$ is the relativity velocity of the incoming DM particles. 
To derive eq.(\ref{ann}) we have used crossing symmetry to transfer the results on squared amplitude $\mathcal{M}\left(XX\rightarrow \bar{\chi}\chi\right)$ calculated by ref.\cite{Clery:2021bwz} to $\mathcal{M}\left(\bar{\chi}\chi\rightarrow XX\right)$ in our case, 
and the non-relativistic approximation on the incoming DM particles which gives $s\approx 4m^{2}_{\chi}$ and $t\approx -m^{2}_{\chi}$.
Plugging the value of $m_{\chi}$ into eq.(\ref{ann}) yields $\sigma\upsilon_{\rm{rel}} \sim 0.1\lambda_{\phi}M^{-2}_{P}\sim 10^{-76}$ cm$^{2}$.
Either neutrino or photon signals with so small strengths and so high energies of order $m_{\chi}$ are beyond the reaches of current neutrino or $\gamma$-ray experiments.

One may alternatively consider DM self interaction via the gravitational portal that leads to the elastic scattering cross section
\begin{eqnarray}
\sigma\left(\bar{\chi}\chi\rightarrow \bar{\chi}\chi\right)\upsilon_{\rm{rel}} \approx  \frac{1}{32\pi}\frac{m^{2}_{\chi}}{M^{4}_{P}},
\end{eqnarray}
which gives $\sigma\left(\bar{\chi}\chi\rightarrow \bar{\chi}\chi\right)/m_{\chi}\sim\lambda^{\frac{1}{2}}_{\phi}M^{-3}_{P}\sim 10^{-60}$cm$^{2}/$g for $\upsilon_{\rm{rel}}\sim 10^{-4}$.
This suggests that non-gravitational DM self interaction is required to obtain $\sigma/m_{\chi}\sim \mathcal{O}(1)$cm$^{2}/$g \cite{Tulin:2017ara} as inferred from the assumption that DM self interaction is the solution to small-scale problem.
Note, there is no obstacle to introduce DM self interaction in our situation, 
as it does not effect the estimate on the DM parameter space in fig.\ref{ps}.

\section{Conclusion}
\label{con}
In this work we have studied the gravitational freeze-in DM production under the context of Higgs preheating.
By taking into account both the cosmological and collider constraints on relevant model parameters, 
we have used the lattice calculation to analyze the non-perturbative process taking place during the preheating.
We have shown that the tachyonic resonance is forbidden by the back reaction due to the Higgs self interaction which is needed to keep the positivity of the potential during preheating, but the parameter resonance is still viable as long as the effects of the back reaction are under control (by tuning the Higgs self coupling constant in the ultraviolet energy region).
Further, we have derived the formula of DM relic density in the DM mass range with $m_{\chi}>m_{\phi}$ arising from the Higgs annihilation during preheating.
Using this formula we have presented the new DM parameter space in terms of numerically solving the Boltzmann equation.
Finally, we have briefly discussed the phenomenology of gravitational DM,  
verifying the weakness of traditional search strategies such as the DM annihilation. 

We believe that at least the following points deserve further investigations. 
$(1)$ Our lattice results about the Higgs preheating can shield light on DM preheating.
They demonstrate that apart from the magnitude of interaction between the inflaton and Higgs sector, 
the Higgs self interaction is another key factor to shape the Higgs preheating. 
This insight can be directly applied to constrain DM preheating if DM self interaction is believed to be the solution to small-scale problem. 
$(2)$ In principle a higher PC source enables us to improve the DM parameter space with the DM mass $m_{\chi}>110~m_{\phi}$ which is beyond the scope of this study.
$(3)$ Testing the gravitational DM via traditional windows seems implausible.
It should be of interest to develop other new directions \cite{XZZ}.

\section*{Acknowledgments}
This research is supported in part by the National Natural Science Foundation of China under Grant No. 11775039.

\end{document}